\documentclass[12pt]{iopart}

\usepackage{graphicx}
\begin{document}

\title{Experimental and computational studies of jamming}

\author{Chaoming Song, Ping Wang, Fabricio Potiguar,
Hern\'an A. Makse \footnote[3]{To
whom correspondence should be addressed (makse@mailaps.org)}
}

\address { Levich Institute and Physics Department\\ City College
of New York\\ New York, NY 10031, US}

\begin{abstract}
Jamming is a common feature of out of equilibrium
systems showing slow relaxation dynamics.
Here we review our efforts in understanding jamming in
granular materials using experiments and computer simulations.
We first obtain an estimation of an effective temperature
for a slowly sheared granular material very close to jamming.
The measurement of the effective temperature is
realized in the laboratory by slowly shearing a closely-packed
ensemble of spherical beads confined by an external pressure in a
Couette geometry. All the probe particles, independent of their
characteristic features, equilibrate at the same temperature, given by
the packing density of the system.  This suggests that the effective
temperature is a state variable for the nearly jammed system.
Then we investigate numerically whether the effective temperature
can be obtained from a flat average over the jammed configuration
at a given energy in the granular packing, as postulated
by the thermodynamic approach to grains.

\end{abstract}



\maketitle

\clearpage

\section{Introduction}

The application of ideas from
equilibrium statistical mechanics to nonequilibrium systems
has been extensively debated in the literature
\cite{liu1,mehta,coniglio,wolf} and
has found examples in
structural glasses \cite{grigera}, colloidal suspensions
\cite{bellon}, spin glasses \cite{herisson}, highly agitated granular
matter \cite{danna} and a fluidized particle \cite{durian}.

It has been suggested that, under certain experimental conditions, a
jammed granular system could be amenable to
a statistical mechanics formulation \cite{edwards}.
This requires new statistical concepts specific
to this kind of a dissipative athermal system.
Provided that all the jammed configurations of the system are equally
probable and bear no memory of their creation, we arrive at the
ergodic hypothesis, implying that a statistical mechanics approach is
justified. This forms the basic tenet of the statistical mechanics
formulation for jammed granular matter, which characterizes the packing state
by the entropy, $S$,  and the compactivity, $X$,
defined as $X=\partial V/\partial S$, where $V$ is the system volume
\cite{edwards,mbe}.
The existence of jammed reversible states in granular matter
has been suggested by
compaction experiments employing tapping, oscillatory compression or
sound propagation as the external perturbation
\cite{nowak,bideau,cavendish,edwards2,bideau2}.

Recent studies indicate that out-of-equilibrium systems
showing slow relaxation dynamics, such as an aging
glass or a slowly sheared granular material close to jamming,
can be characterized by an ``effective temperature''
arising from an extension of
the fluctuation-dissipation
relation to nonequilibrium situations \cite{ckp}.
Whereas the theoretical concept of compactivity
has been developed for static jammed configurations of grains,
the analogy with dynamical measurements of effective temperatures may
exist if the motion is sufficiently slow.
In fact, the resulting effective temperature
is closely related to the compactivity $X$,
as obtained by a flat average
over the available jammed configurations.
Until present, the only evidence for the
significance of these ideas
in describing granular matter has emerged from theoretical
mean-field models \cite{kurchan} and computer
simulations of glassy systems and
granular matter \cite{liu3,bklm,ono,coniglio2,mk}.
Macroscopic
variables,
 such as the effective temperature for granular matter,
have not been previously
measured in the laboratory.
This line of research has led
to the design of the experiment which we are about to describe.

In this paper we first present experimental evidence \cite{swm}
for the validity of the
effective temperature in describing a slowly sheared granular material
very close to jamming.
The particle
trajectories in the sheared system
yield the diffusivity and the mobility from which the
temperature is deduced using a fluctuation-dissipation
relation.  All the particles equilibrate at the same temperature,
which is in turn
independent of the slow shear rate, thus suggesting the condition of a
state variable for the jammed system.
In the second part of this paper we review our
efforts \cite{mk,potiguar}
to understand the meaning of the effective temperature
from a thermodynamic point of view under the framework
of the statistical mechanical approach introduced by Edwards \cite{edwards}.

\section{Experiments}

\subsection{Experimental set-up}

\begin{figure}
\centering
{(a) \resizebox{12cm}{!}{\includegraphics{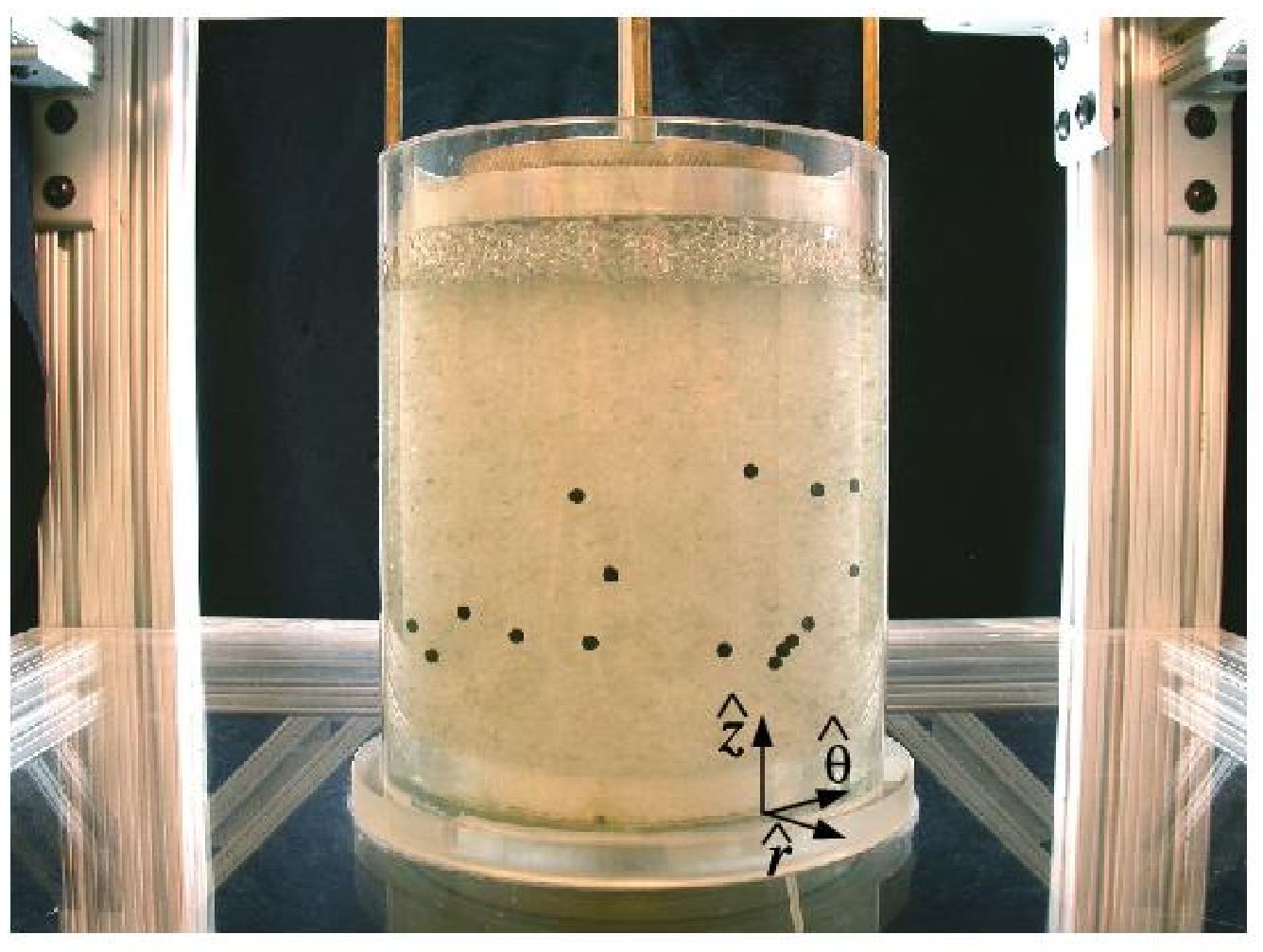}}}

\centering{
(b) \resizebox{8cm}{!}{\includegraphics{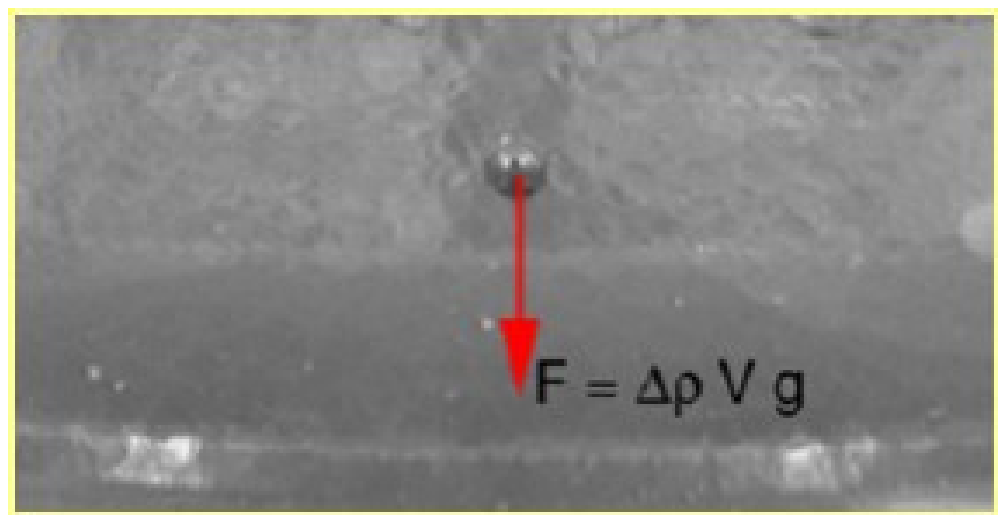}}
}
  \caption{Experimental set-up.
(a) Transparent acrylic grains and black tracers
in a refractive index and density matched solution
are confined between the inner cylinder of radius $50.8$ mm
and the outer cylinder of radius $66.7$ mm.
(b) Detail of a metallic tracer in the background of PMMA particles.}
\label{couette}
\end{figure}

The experimental test for the application of
statistical mechanics concepts to jammed matter involves using
observational techniques to monitor the evolution of the particulate
packing as it explores the available
configurations.
 The different packing configurations are investigated using
quasi-static shear in a Couette cell geometry, depicted in Fig.
\ref{couette}.

The particulate system is confined between an inner and an outer
cylinder. The inner cylinder of the cell is slowly rotated by a motor
while the outer cylinder is fixed and transparent for
visualization. The walls of both cylinders are roughened by gluing a
layer of
particles to provide shear motion to the assembly, avoiding
wall-slip. The grains are compactified by the application of an
external pressure of a specific value (typically 386 Pa), introduced
by a moving piston at the top of the granular matrix. We use a narrow
gap Couette cell of the order of 7 particle diameters wide in order
to avoid the formation of bulk shear bands
\cite{nedderman,drake,mueth,veje,utter}.

Since a granular assembly is optically impenetrable by its very
nature the first task consists in creating a transparent sample.
This is achieved by refractive index matching transparent particles
with a suitable suspending solution. The presence of the solution
reduces friction between the particles, nevertheless
the ensemble remains jammed
throughout the experiment by the  application of the external force via the
piston.  It is important to note that the liquid only partially fills
the cell (see Figure \ref{couette}).
In this way the pressure of the piston is transmitted only to the
granular particles and not to the fluid.
The key feature of the system
is being closely packed, which is hereby satisfied.
The random motion of the particles is due to the `jamming' forces
exerted by the enduring contacts of all the neighboring particles
which renders the problem nontrivial.

The successful packing consists of a 1:1 mixture of two different
sizes of spherical Poly-methyl methacrylate (acrylic) particles, of
density $\rho = 1.19$ and index of refraction $n=1.49$. We use two
different packings, containing either 3.17 mm and 3.97 mm diameter
particles (Packing 1) or particles of diameter 3.97 mm and 4.76 mm
(Packing 2). The size ratio ensures that crystallization is
avoided. The mixture of particles is immersed in a solution of
approximately 74\% weight fraction of cyclohexyl bromide and 26\%
decalin \cite{weeks}, matching not only the refractive index but more
importantly the density of the acrylic particles. The density matching
fluid eliminates pressure gradients associated with gravity in the
vertical direction. This step avoids problems encountered in previous
tests of compactivity \cite{nowak} and other effects such as
convection and size segregation such as the Brazil nut effect inside
the cell \cite{behringer}.

We follow the trajectories of tracer particles in the sample bulk
to obtain
the diffusion of the tracers and the
response function (mobility) to an external force within the
structure. These measurements lead to the
compactivity or
effective
temperature via a fluctuation-dissipation relation generalized to
granular media.  If the resulting effective temperature is
a physical variable of the system,
it will be
independent of the
properties of the tracer particles-- a necessary but not sufficient condition.
It is to this end that we
contribute experimental results.
We note that to evaluate the broader thermodynamic meaning of the
temperature
it would be needed to examine
whether different measures of temperature agree as well.
Such tests could be performed, for instance, by measuring the temperature
for different observables in the system.

The tracer particles must experience a constant force, in response to
which the mobility can be measured. Tracers of a different density to
the acrylic particles are then added to the packing, as shown in
Fig. \ref{couette}.
Two types of tracers, made of nylon ($\rho = 1.12$) and delrin ($\rho
= 1.36$), are employed.
The role of the tracer in the system is to explore the different
packing configurations, and
the size of the tracers is chosen, accordingly,
of similar size as the  background particles.
If the tracers are much larger or smaller than the background particles
new physics would be brought into the problem.
For instance if a tracer is small
enough to fall into the voids of the other particles
then ``percolation effects'' \cite{drahun}
would prevail and the displacements
would be larger than those predicted by the effective
temperature.
Such a tracer would no longer prove
the background particle network, but would
instead have different dynamics and
test other interesting effects which
cannot be captured by the present formalism.

\subsection{Experimental Results}

The first experiment uses Packing 1 with 20 tracers of  3.97
mm nylon
beads. The Couette cell is sheared at very slow frequencies
$f = 2.4$ mHz defining the external shear-rate $\dot \gamma_e =
2\pi f r_1/(r_2-r_1)= 0.048$
1/s, where $r_1=50.8$ mm and $r_2=66.7$ mm are the radii of the inner
and outer cylinders, respectively.
We require a very slow shear rate so that the system is
close-packed
at all times.
The $(r(t),\theta(t),z(t))$ coordinates of the tracers are obtained by
analyzing the images acquired by four digital
cameras surrounding the shear
cell [($r,\theta$) are obtained
only at the overlaps of the cameras].
In the following we first present results for the $z$ direction since
this is the only direction where the temperature can be calculated
with the present set-up
(the external force acts only vertically).

\begin{figure}
\centerline {\resizebox{10.cm}{!}{\includegraphics{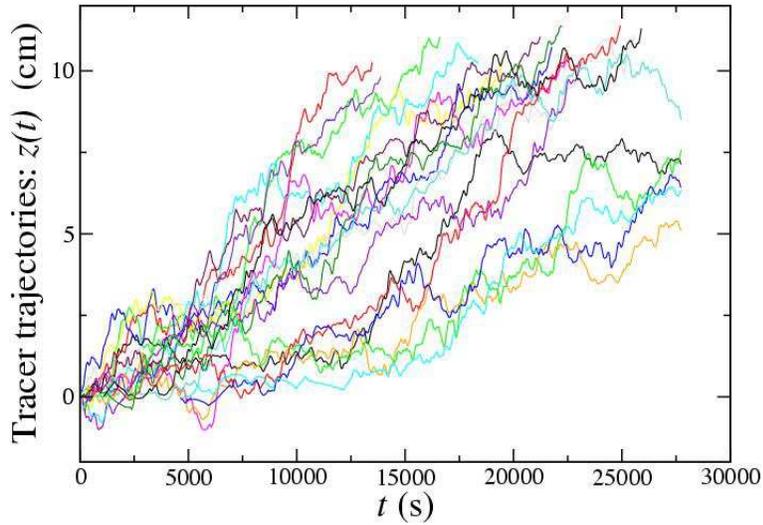}}}
\caption{Trajectories of the 3.97 mm nylon tracers in Packing 1
showing the diffusion and response to
the gravitational force when sheared in the Couette cell.
Note that the tracers diffuse distances larger
than the particle diameter indicating that we are probing fluctuations
outside the ``cages'' formed by the surrounding particles.
}
\label{trajectories}
\end{figure}

The resulting vertical trajectories of the tracers $z(t)$ are
depicted in Fig.  \ref{trajectories} showing that the nylon tracers not only
diffuse, but also move with a constant average velocity to the top of
the cell.
We confine the measurements of the tracer fluctuations
away from the inner rotating cylinder
to avoid boundary effects and where
the average tangential velocity of the tracers,
$v_{\theta}(r)$, can be
approximated linearly
as $v_{\theta}(r)\approx-\dot\gamma_l r$ with
a constant local shear rate $\dot\gamma_l = 0.021$ 1/s. This
ensures that the diffusivity (which depends on the local shear rate)
remains approximately constant in the radial direction.

\begin{figure}
\vspace{1cm}
\centering {\resizebox{10cm}{!}{\includegraphics{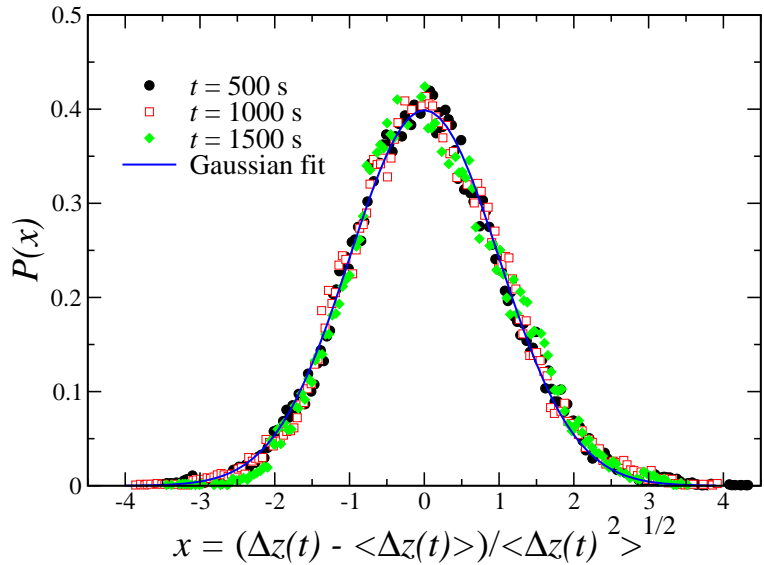}}}
\caption{Probability distribution of the displacements of the
3.97 mm nylon tracers in Packing 1.
The different distributions at different times are
rescaled according to a Gaussian distribution.
}
\label{gauss}
\end{figure}

\begin{figure}
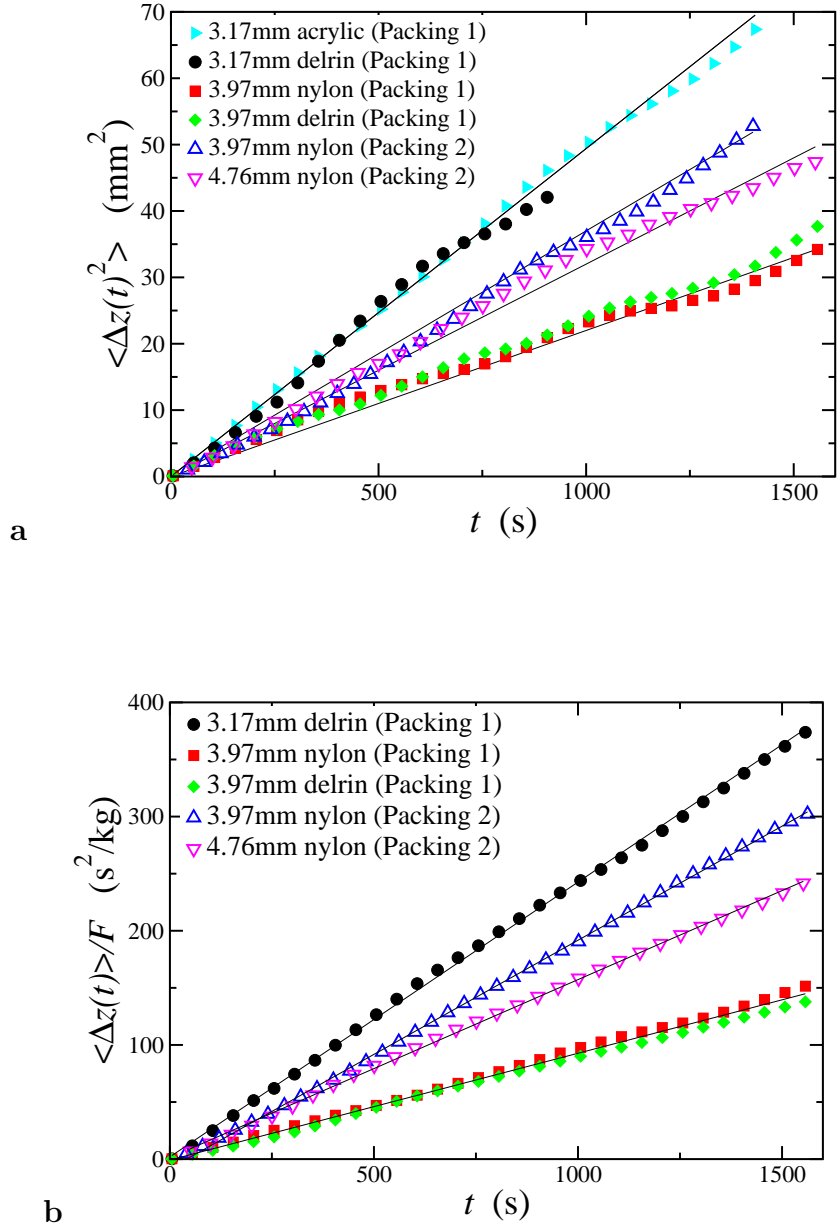

\centering{{\bf a} \hspace{.5cm}
\resizebox{10cm}{!}{\includegraphics{makse-fig4a.eps}}
}
\vspace{2cm}

\centering{ {\bf b}  \resizebox{10cm}{!}{\includegraphics{makse-fig4b.eps}}}
\caption{(a) Diffusion and (b)
mobility of tracers.  We use
Packing 1 and Packing 2 of
acrylic particles and tracers of different sizes and densities.
Packing 2 is run at $\dot\gamma_e = 0.024$ 1/s.
For both packings,  $D$ and $\chi$ are inversely
related to the tracer sizes.
}
\label{mobility}
\end{figure}

The statistical analysis of the particle displacements $\Delta z(t) =
z(t+t_0) - z(t_0)$
%
reveals a Gaussian distribution
which broadens with time, as seen in Fig. \ref{gauss}.  The rms
fluctuations grow linearly for sufficiently long times
(see Fig. \ref{mobility}a):
\begin{equation}
\langle[z(t+t_0) - z(t_0)]^2\rangle \sim 2 D t,
\end{equation}
where $D$ is the self-diffusion constant and $\langle\cdots\rangle$
denotes ensemble average over the tracers and over the initial time $t_0$.
For the 3.97 mm tracer we obtain
$D_{\mbox{\scriptsize 3.97mm}} = ( 1.1 \pm 0.1 )\times 10^{-8}$ m$^2$/s.

Figure \ref{mobility}b shows the mean value of the position of the
tracers extracted from the peak of the Gaussian distribution as a
function of time, thus yielding the mobility $\chi$ as
\begin{equation}
\langle z(t+t_0) - z(t_0)\rangle \sim \chi F t.
\end{equation}
Here $F = (\rho_{a} - \rho_{t}) V g$ is the gravitational force
applied to the tracers due to their density mismatch while $\rho_{a}$
and $\rho_{t}$ are the densities of the acrylic particles and the
tracers respectively, $V$ is the volume of the tracer particle and $g$
the acceleration of gravity.
The value of the mobility for the 3.97mm tracer is
$\chi_{\mbox{\scriptsize 3.97mm}} = ( 9.7 \pm 0.9)
 \times 10^{-2} $ s/kg.
We check that the mobility of the tracers
is constant in the region of measurements.

An important task is to determine whether there exists a linear
response regime in
the system,
which would imply that the mobility
is independent of the external gravitational force as $F\to 0$.  The external
force is varied by changing the density of the tracers of the same
size, under the assumption that the surface properties remain
unchanged. This is realized experimentally in Packing 1
by the introduction of
delrin tracers of 3.97 mm diameter, the density of which is higher than
that of nylon. The analysis of the trajectories
reveals that the  mobility is
the same for both tracers so that it is
independent of the external force, as
shown in Fig.  \ref{mobility}b. The use of tracers even heavier than
delrin reveals the
appearance of nonlinear effects in the mobility.
Moreover we show that the
diffusion constants of both types of tracers
are approximately the same (see Fig. \ref{mobility}a) confirming that
the external force on the tracers does not affect the diffusion
constant for the small forces used in this study.

\subsection{Effective temperature}

According to a Fluctuation-Dissipation relation, it is the diffusivity
and the mobility of the particles which enable the calculation of an
effective temperature, $T_{\mbox{\scriptsize eff}}$, via an Einstein
relation for sheared granular matter:

\begin{equation}
\langle [z(t+t_0) - z(t_0)]^2\rangle = 2 T_{\mbox{\scriptsize eff}}
\frac{\langle z(t+t_0) - z(t_0)\rangle }{F}.
\label{fdt}
\end{equation}

\begin{figure}
\vspace{1cm} \centering {
\resizebox{10cm}{!}{\includegraphics{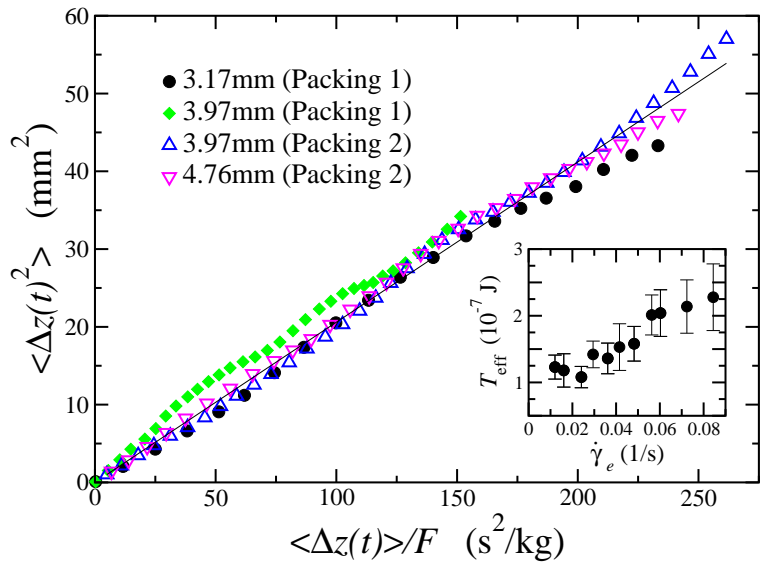}}}
\caption{Effective temperatures for various tracers and different
packings as obtained from a parametric plot of their diffusion versus
mobility, as explained in the text. The slopes of the curves for
different tracers consistently yield the same average value of
 $T_{\mbox{\scriptsize eff}} = (1.1 \pm 0.1) \times 10^{-7}$ J
as given by Eq. (\protect\ref{fdt}).
The inset shows the dependence of
$T_{\mbox{\scriptsize eff}}$
on the shear rate $\dot\gamma_e$ for the 4.76 mm nylon tracers in Packing 2.
We find that $D\sim \dot\gamma_e$ and $\chi\sim \dot\gamma_e$,
while $T_{\mbox{\scriptsize eff}}=D/\chi$ is shear-rate
independent
for sufficiently small  $\dot\gamma_e$.
}
 \label{teff}
\end{figure}

A parametric plot, with $t$ as a parameter, of the fluctuations and
responses is produced to yield the linear relationship shown in
Fig. \ref{teff}, the gradient of which gives $2 T_{\mbox{\scriptsize eff}}$.
We obtain for
the 3.97 mm tracer
 $T_{\mbox{\scriptsize eff}} = (1.1 \pm 0.1) \times 10^{-7}$ J.
This value is set by a typical energy scale
in the system  \cite{review}, for instance
$(\rho_{a} - \rho_{t}) g d$, which is the gravitational
potential energy to move a tracer particle
a  distance of the particle
diameter $d$.
The corresponding temperature
which would arise from the conversion of this energy into a
temperature via the Boltzmann constant, $k_B$, is
$T_{\mbox{\scriptsize eff}}
= 2.7 \times 10^{13} k_B T$ at room temperature. This large value is
expected \cite{review} (and agrees with
computer simulation estimates \cite{mk}) since granular matter is an
athermal system.
We notice that previous results  \cite{mk} suggest that Eq. (\ref{fdt}) is
equivalent to the average over jammed configurations, then  Eq.
(\ref{fdt}) can be used to obtain an estimate of
the compactivity of the packing
\cite{edwards}
as $X=T_{\mbox{\scriptsize eff}}$.

An important evidence in examining the thermodynamic meaning of the
effective temperature can be obtained from a test of the zeroth law.
In other words, changing the tracer size should give rise to a different
diffusion and mobility but they should nevertheless
lead to the measurement of the same
effective temperature, if
the system is at ``equilibrium''.
We next introduce tracers of 3.17 mm diameter in
Packing 1 and repeat the above calculations.  We find that the 3.17 mm
tracers produce a significantly different diffusion and mobility than
their 3.97 mm counterparts as shown in Fig. \ref{mobility}
($D_{\mbox{\scriptsize 3.17mm}} = ( 2.5 \pm 0.3) \times 10^{-8}$ m$^2$/s
and $\chi_{\mbox{\scriptsize 3.17mm}} = ( 2.4 \pm 0.3) \times 10^{-1}$ s/kg).
In all
cases $D$ and $\chi$ increase with decreasing size of the tracers.
However, the parametric plot of  diffusivity versus mobility
demonstrates that their effective temperature are the same as seen in
Fig.  \ref{teff} with an average value over all tracers of
 $T_{\mbox{\scriptsize eff}} = (1.1 \pm 0.1) \times 10^{-7}$ J.

We further check that the diffusion is not affected by the
external force by calculating the diffusivity of the nontracers
particles by dying acrylic tracers and analyzing their
trajectories.
As shown in Fig. \ref{mobility}a
the diffusion of the acrylic tracers of size
 3.17 mm (for which no external force is applied)
is the same as the diffusion of the delrin tracers of the same size
(for which the gravitational force
is applied).

Next we perform
another
consistent
measurement arising from a repeat of the experiment for a
different packing of spherical particle (Packing 2).
The use of larger particles of
approximately the same size ratio as in Packing 1 still leads to
the same volume fraction of
particles.
Since  $T_{\mbox{\scriptsize eff}}$ is a
measure of how
dense the particulate packing is (i.e. a large  $T_{\mbox{\scriptsize eff}}$
implies a
loose configuration, e.g. random loose packing, while a reduced
$T_{\mbox{\scriptsize eff}}$ implies a more compact structure, e.g.
random close
packing),
it holds to reason that it should be the same for both of the packings
under investigation.
Indeed, despite the change in their
respective diffusivities and mobilities as shown in Fig.
\ref{mobility}, the two packings measure the same effective
temperature shown in Fig. \ref{teff}.

An important assumption in this study is that of the system being
continuously jammed despite the presence of rearrangements under
shear. For this reason, we show in the inset of Fig. \ref{teff} that
the effective temperature is approximately
independent of the shear rate, as long as
the particulate motion is slow enough such that enduring contacts
prevail. We find that $D\sim \dot\gamma_e$ and $\chi\sim \dot\gamma_e$,
while $T_{\mbox{\scriptsize eff}}=D/\chi$ remains approximately constant
for sufficiently small  $\dot\gamma_e$.
It is within this quasi-static range where the
 effective temperature could be identified
with the temperature of the jammed states.
This quasi-static regime has been obtained in previous computer simulations
\cite{ono}, and it is shown in the numerical section of this paper.
 The regime corresponds to the regime where the stress in the
system becomes shear-rate independent \cite{tardos}.

Given that we are dealing with an athermal system in which the notion
of 'bath' temperature
plays no role, but
the thermodynamic
concept of temperature still holds, perhaps
further explanation is required in terms of the actual role of
 $T_{\mbox{\scriptsize eff}}$ in describing granular systems.
The length scale on which the particles diffuse over the long time
scale of the experiment is of the order of several particle diameters
(see Fig. \ref{trajectories} and \ref{mobility}a)
implying that the exploration of the available
jammed  configurations
takes place by rearrangements of the particles outside their ``cages''.
The trajectory of the slow moving system
can be mapped onto the successive jammed states that the system
explores.
This paints a configurational landscape of jamming,
analogous to that observed in the inherent structures formalism of glassy
materials \cite{debenedetti,coniglio}.
Thus,  we identify
$T_{\mbox{\scriptsize eff}}$ as the variable governing
this exploration of the
different jammed configurations.
In
contrast to the measurement of the temperature of the slow modes,
we also measure the temperature of the fast modes as
given by the instantaneous rms fluctuations of the
velocity of the particles inside each
cage. This kinetic granular temperature is smaller than
$T_{\mbox{\scriptsize eff}}$ and
differs for each tracer indicating that it is not governed by the same
statistics. Similar results have been obtained in experiments
of vibrated granular gases \cite{menon}.

\subsection{Outlook}

In the next section we present our computational efforts to understand jamming.
We will treat the question whether the
present definition
of effective temperature can be related
to the thermodynamic formalism of grains.
The main assumption of this statistical formulation
is that the  different jammed configurations are taken to have
the same statistical weight \cite{edwards}. Thus,
observables can be obtained as ``flat'' averages of the jammed
configurations \cite{kurchan,bklm,coniglio2,mk}.
The  validity of this assumption
has been extensively
debated in the literature (see for instance \cite{coniglio}).
Some simulations and analytical work
suggest that the effective temperature obtained by applying
the extension of FDT to out-of-equilibrium systems
is indeed analogous to performing a flat average over
the configurational space.
Numerically it has been suggested that the
effective temperature can be identified with the
compactivity introduced in \cite{edwards},
arising from the entropy of the packing \cite{kurchan,bklm,mk}.
In the next section we present our studies showing that
under some conditions the effective temperature can be related to
a temperature obtained from entropic considerations
using a flat average over the jammed configurations.
We review the results published in \cite{mk,potiguar}.

\section{Numerical studies}

This  section describes the potential for using computer
simulations in testing the thermodynamic foundations raised in the
previous sections. Rather than employing rough rigid grains for
which most of the theoretical concepts have so far been devised,
computer simulations are obliged to introduce some deformability
into the constituent particles to facilitate the measurement of
the particle interactions with respect to their positions. As a
result, the entropic considerations which have been explained only
in terms of the volume in the original formulation of
Edwards \cite{edwards} for the
case of rigid grains will now be generalised to situations in which there
is a finite energy of deformation in the system \cite{mbe}.

The entropy of the jammed packing can be redefined as a
function of both energy and volume,
\begin{equation}
S(E,V) = \lambda \ln \Sigma_{\mbox{\scriptsize jammed}}(E,V),
\end{equation}
where  $\Sigma_{\mbox{\scriptsize jammed}}(E,V)$ measures
the number of microstates for a given volume $V$ and energy $E$.
The introduction of energy into the system implies a corresponding
compactivity,

\begin{equation}\label{compactivityE}
X_E^{-1} = \frac{{\partial S} }{{\partial E}},
\end{equation}
where the subscript $E$ denotes that the compactivity is  the
Lagrange multiplier controlling the energy of the jammed
configuration, not the volume. Notice that $X_E$ differs from the
temperature of an equilibrium system $T = \partial E/\partial S$
because the energy in Eq. (\ref{compactivityE}) is the energy of
the jammed configurations and not the thermal equilibrium energy.
The radical
step is the assumption of equally probable microstates
which leads to an
analogous thermodynamic entropy associated with this statistical
quantity.

In the next sections we present results for the effective temperature, defined through the
 fluctuation-dissipation theorem, of granular material obtained from extensive computer
simulations of a sheared granular packing of compressible spheres. We show that this
parameter is independent of the shear rate in a defined interval for high volume
 fractions, and does not depend on the particular species of particle where it is measured.
We also compare these measurements with
the granular temperature and show that, indeed, the last one is not the thermodynamical
 parameter for the jammed states. We then show that the temperature can be obtained as a flat
average over the jammed configurations.
This approach was  validated numerically in \cite{mk}
 who showed that the temperature obtained by performing a flat
average over the jammed configuration
is the same, within error bars, as the effective temperature
obtained dynamically from a fluctuation-dissipation
 relation in a gently sheared, bi-disperse assembly of granular
 balls.

\subsection{The system}

We used Molecular Dynamics \cite{Cun79} to model our
system. This system is composed of a bi-disperse assembly of 1000
spheres, half large, half small, with radii ratio of $R_S/R_L\sim
0.818$, which are initially randomly generated in a periodic cubic
cell of size $L$.
The spheres interact via normal and tangential forces. To
describe these interactions we consider two spheres in contact
following the contact force model of Hertz-Mindlin \cite{Mak00}.
The normal force is given by the Hertz model such that the
force is proportional to the  deformation of the spheres to the 3/2.
The
tangential force between two spheres in contact is given by Mindlin
no-slip solution for the contact. This force is proportional to the
magnitude of the normal force and depends on the
loading path of contact.
During the contact, the tangential force
between two grains can increase until it reaches the Coulomb threshold
of static friction, $F_t=\mu F_n$
 in which case the grains slide.
Gravity is absent in all our simulations. We can also study
packings in which the particles interact only by normal forces, and
this case will be referred to as the no-friction, or frictionless,
case. We notice that the frictionless case would model a system of
compressed emulsions \cite{mbe}.

Since our goal is to study dense systems, we should compress the
initial packing in order to attain a higher volume fraction. It is
known \cite{Mak00,ohern}
that there is a jamming transition at a critical volume
fraction $\phi_c$ of randon close packing (RCP)
 above which a disordered solid state persists indefinitely. This
metastable state is characterized by a non-vanishing internal pressure
and coordination number. Therefore, we choose three initial
equilibration pressure values, $P=1,10,100$ MPa, that are high
enough to generate a disordered jammed state. The corresponding initial
volume fractions of these states are $\phi = 0.6428, 0.6623, 0.7251$,
above the RCP volume fraction. We also consider systems below the jamming
transition
in the ``fluid-like'' state
for volume fractions $\phi = 0.6318, 0.6104, 0.5999 < \phi_c$.

\begin{figure}
\centering{
 \resizebox{10cm}{!}{\includegraphics{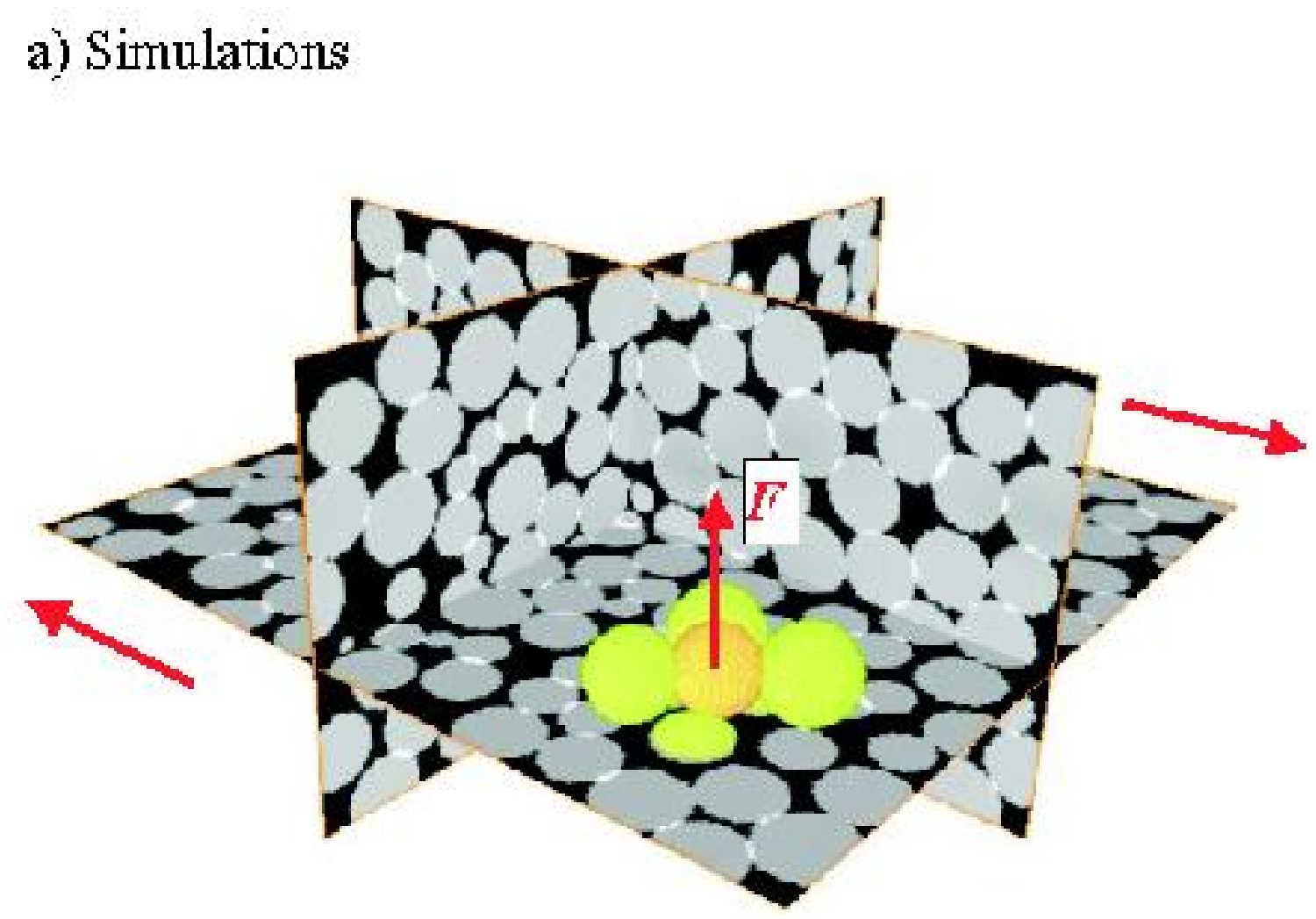}}
}
\caption{Detail of the simulations of grains of 100 microns interacting via
Hertz-Mindlin contact forces. A slow shear flow, indicated by the
arrows, is applied to the jammed system. We follow the tracer
particle trajectories to obtain the diffusivity. An external force
$F$ is then applied to the tracers in response to which we measure
the particle mobility. These dynamical measurements yield an
effective temperature obtained from an Einstein relation.}
\label{amira}
\end{figure}

We consider a system of large and small spherical grains in a
periodic cell. The simulations involve the application of a gentle
shear on the particles in the $x$-$y$ plane, at constant volume
(see Fig. \ref{amira}), with $x$ the direction of the flow and
$y$ the velocity gradient. We follow the trajectory of the particles
in the $z$-vorticity direction.
We used a modified version of the
usual Lees-Edwards boundary conditions  which imposes a
linear velocity profile in the shear plane.
Normal periodic boundary conditions are enforced in the
$z-$direction. We focus our
study on the region of slow shear rates, where the system is
always close to jamming. We avoid shear bands by imposing a linear
velocity profile and avoid segregation which may occur at much
longer time scales than those employed in our computations.

\subsection{Diffusion, mobility and effective temperature}

We calculate the effective temperature, $T_{\mbox{\scriptsize eff}}$, of the granular system
 by measuring it
through the same FDT as in the experiments, via diffusion and mobility.

The calculation of the diffusion coefficient involves the  square
displacement averaged over all particles and over time. All quantities
which depend on the Cartesian coordinates will be evaluated in the
direction perpendicular to the shear plane, in our case $z$-direction,
in order to avoid complications with shear-induced displacements.
The mobility of a particle, $\chi$, is the linear response function
associated with the motion of this particle under an applied
external force. To calculate $\chi$,
a small constant force must be applied to the particle in order to
induce a displacement in $z$-direction.
The effective temperature is defined from the  Einstein relation between
$D$ and
$\chi$ as in Eq. (\ref{fdt}).

\begin{figure}
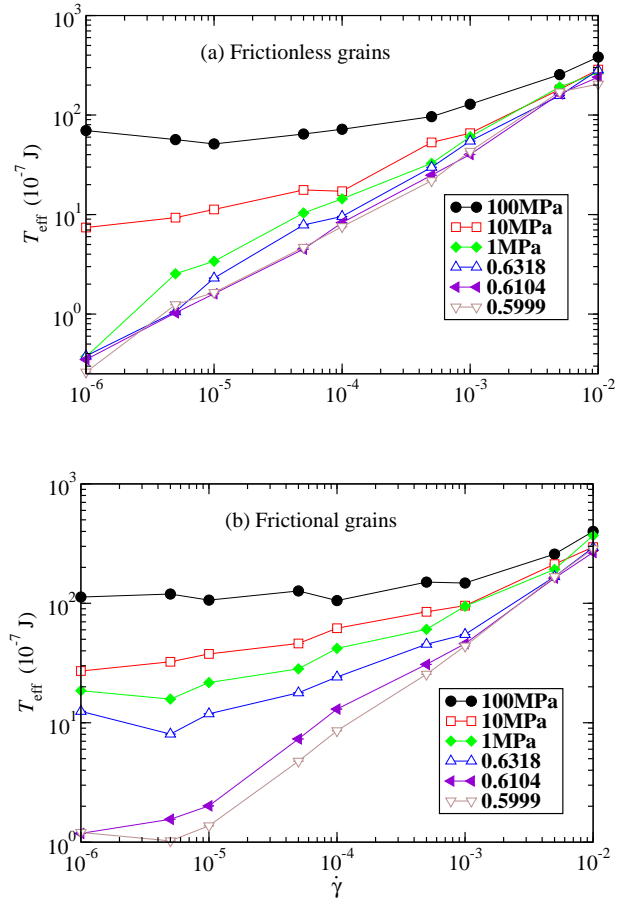

\centering
{\resizebox{8cm}{!}{\includegraphics{t-shear-no-fric.eps}}}

\vspace{.5cm}

\centering{
\resizebox{8cm}{!}{\includegraphics{t-shear-fric.eps}}
}
\caption{
Effective temperature versus shear rate for (a) frictionless particles
and for (b) frictional particles. We consider systems above the
jamming transition for pressures $p=1$ MPa, 10 MPa, and 100MPa,
and below the jamming transition for volume fractions
$\phi=0.6318, 0.6104, $ and 0.5999.
For the frictionless case, the independence of
$T_{\mbox{\scriptsize eff}}$ with the
shear rate is only observed in the 10 and 100 MPa cases.
The other states seem to be unstable under shear.
For frictional particles,
the effective temperature is approximately independent with the shear
rate even in the $\phi=0.6318$ case. This is evidence that the jammed state
could be attained by a granular system without any initial
compression.}
\label{teff-fric}
\end{figure}

We plot in Fig.
 \ref{teff-fric},  $T_{\mbox{\scriptsize eff}}$ as a
function of the shear rate $\dot\gamma$ for
different pressures and volume fractions.
 We can see that at higher volume
fractions, in no-friction and friction cases, this effective
temperature is approximately independent on the shear rate. This is an
important result for it shows that this is indeed an intrinsic property of
the jammed system, since it does not depend on the driving strength. The
jammed state for $\phi>\phi_c$
can be characterized by the independence of the effective
temperature on the shear rate since this behaviour is not observed in
the other cases studied.

\begin{figure}[!tbp]
\centering
{ \resizebox{12cm}{!}{\includegraphics{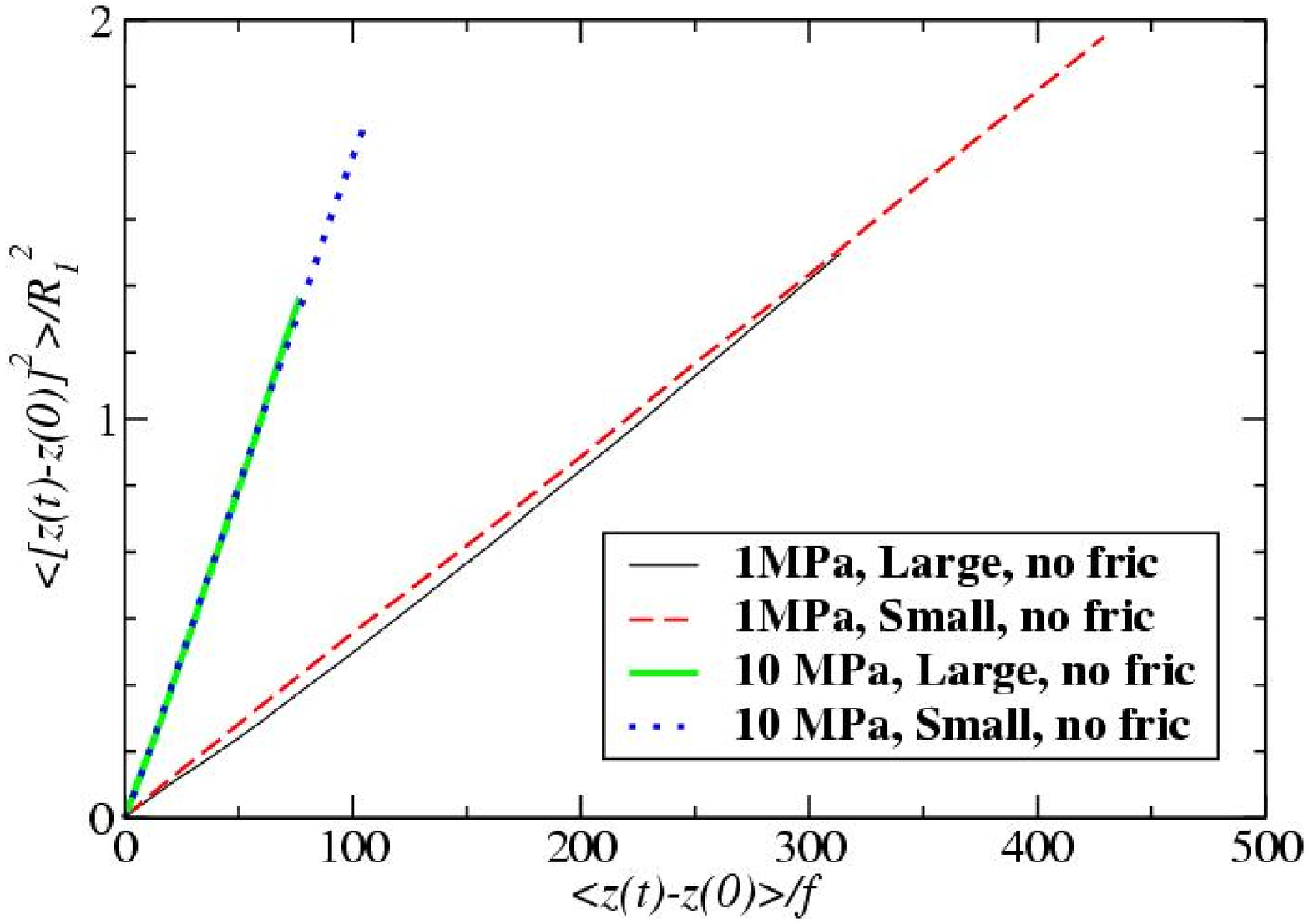}}}
\caption{Parametric plot of diffusion coefficient versus mobility. The
equality of the effective temperatures for both species of particles
is evident in this plot. This gives another evidence that $T_{\mbox{\scriptsize eff}}$ is
an intrinsic property of the system. The legend displays the
equilibrated pressure, the size of the particles and if they have
friction or not. The curves for the $10$ MPa case overlap almost perfectly.}
\label{para-plot}
\end{figure}

Next we perform a similar test of the zeroth-law as done experimentally.
We test this by repeating the mobility and diffusion
calculations with the small particles for
two system of frictionless grains at 1 MPa and 10 MPa.
Figure \ref{para-plot} shows the plot of diffusion versus mobility for the
large and small frictionless particles at 1 and 10 MPa, sheared at
$\dot\gamma = 5\times10^{-6}$. The agreement is clearly excellent.
This further suggests that $T_{\mbox{\scriptsize eff}}$ is a good thermodynamic
variable for the system.

Finally, we compare the effective temperature with the kinetic
granular temperature
as done in the experiments.
As in thermal systems, we can define a parameter called kinetic
granular temperature, $T_k$, which is the
average of the kinetic energy. This is an estimate of the
temperature of the fast
modes of the system (since a granular system is athermal by
definition) and, in general, should
not be equal to the effective temperature.

Our numerical results confirm that $T_{\mbox{\scriptsize eff}}$ is
 always larger than the granular temperature, $T_k$. We also obtain
 that the behavior of $T_{\mbox{\scriptsize eff}}$ with volume
 fraction is different from the behaviour of $T_k$ in $\phi$. The last
 one decreases with increasing $\phi$ while the former increases. This
 implies that the effective temperature is the temperature of the slow
 modes, while the granular temperature is the temperature of the fast
 modes of our system, and the last one is not the temperature for the
 jammed state.

The next crucial test for this assumption is to show that the effective
temperature obtained dynamically can also be obtained via a flat
average over the jammed configurations. Such a test has been
performed in \cite{mk}, where it was indeed shown that
$T_{\mbox{\scriptsize eff}}$ is very close to the compactivity of
the packing $X_E$. This result will be shown explicitly in the next section.
We conclude that the jammed configurations
explored during shear are sampled in an equiprobable way as
required by the ergodic principle. Moreover the dynamical measurement of
compactivity renders the thermodynamic approach amenable to
experimental investigations as done in the first part of this review.

In the next section we show the relation between
 the effective temperature of the
packing obtained dynamically and the compactivity
calculated employing a configurational average.

\subsection{Exploring the jammed configurations via a flat average.
Test of ergodicity: $T_{\mbox{\scriptsize eff}} = X_E$}
\label{ergodicity}

The systems under investigation have exponentially large (in the
number of particles) number of stable states jammed at zero bath
temperature. In the previous section we explored such an energy
landscape via slow shear. Next, we  develop an independent method
to study the configurational space. It  allows us to investigate
the statistical properties of the jammed states available at a
given energy and volume. In turn we  investigate whether it is
possible to relate the dynamical temperature obtained above via a
diffusion-mobility protocol to the configurational compactivity
based on a flat average over jammed states.

In order to calculate $X_{\mbox{\scriptsize E}}$ and compare with
the obtained $T_{\mbox{\scriptsize eff}}$ we need to sample the
jammed configurations at a given energy and volume in a
{\it equiprobable} way. In order to do this we sample the jammed
configurations with the following probability distribution:
\begin{equation}
P_\nu \sim \exp[- E^\nu/T^* - E^\nu_{\mbox{\scriptsize
jammed}}/T_{\mbox{\scriptsize aux}}] \label{partition}
\end{equation}
We consider only frictionless grains for this calculation, so
the deformation energy $E$ corresponds to the  Hertzian
energy of deformation of the grains. The extra term added in
Eq. (\ref{partition})
 allows us to perform the flat sampling
of the jammed states.
 The jammed energy is such that it vanishes
at the jammed configurations:
\begin{equation}
E_{\mbox{\scriptsize jammed}} \propto \sum_a
\left|\vec{F}_a\right|^2,
\end{equation}
where $\vec{F}_a$ is the total force exerted on particle $a$ by
its neighbours.

\begin{figure}
\centering{
 \resizebox{10cm}{!}{\includegraphics{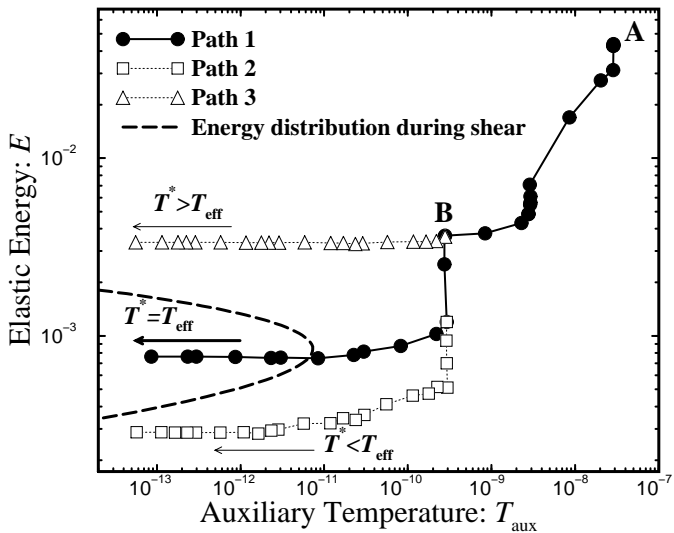}}
}
\caption{ Annealing procedure to calculate  $X_E$ at different
elastic compressional energies. For this calculation we use a system
of 200 frictionless particles. The small system size is needed due to the intensive calculations required by the annealing procedure.
 We plot the elastic energy vs
$T_{\mbox{\scriptsize aux}}$ during  annealing together with
the distribution of elastic energies obtained during shear (dashed
curve, mean value $\langle E \rangle = 8.4\times 10^{-4}$). We
equilibrate the system for $40\times 10^6$ iterations at A:
$(T^*=3.4\times 10^{-2}, T_{\mbox{\scriptsize aux}}=3\times
10^{-8})$. We then anneal slowly both temperatures until B:
$(T^*=3.4\times 10^{-4}, T_{\mbox{\scriptsize aux}}=3\times
10^{-10})$, where we split the trajectory in three paths in the
$(T^*, T_{\mbox{\scriptsize aux}})$ plane.  Path 1: we anneal $
T_{\mbox{\scriptsize aux}}\to 0$ and $T^* \to 2.8\times 10^{-5}$
which corresponds to $ T_{\mbox{\scriptsize eff}}$ obtained during
shear.  Path 2: we anneal $ T_{\mbox{\scriptsize aux}}\to
0$ and $T^* \to 3.4\times 10^{-6}$. Path 3: we anneal $
T_{\mbox{\scriptsize aux}}\to 0$ but keep $T^*=3.4\times 10^{-4}$
constant.  When we set $T^* = T_{\mbox{\scriptsize eff}}$ (Path
1), the final elastic compressional energy value when $
T_{\mbox{\scriptsize aux}}\to 0$ is very close to the mean value of the
elastic energy obtained during shear $\langle E \rangle$.
The value of the mean energy during shear is
 obtained for the same system as used in the annealing calculations.
This demonstrates that
$ T_{\mbox{\scriptsize eff}} = X_E$ under
the numerical accuracy of the simulations.  For other values of
$T^* \ne T_{\mbox{\scriptsize eff}}$ the final $E$ falls out of
the distribution obtained during shear (Paths 2 and 3).  We also
follow different trajectories (not shown in the figure) to $T^*
\to 2.8\times 10^{-5}, T_{\mbox{\scriptsize aux}}\to 0$ and find
the same results indicating that our procedure is independent of
the annealing path. } \label{annealing}
\end{figure}

We introduce two ``bath'' temperatures which will allow us to
explore the configuration space and calculate the entropy of the
packing assuming a flat average over the jammed configurations. We
perform equilibrium MD simulations with two
auxiliary ``bath'' temperatures $(T^*,T_{\mbox{\scriptsize
aux}})$, corresponding to the partition function
(\ref{partition}). Annealing $T_{\mbox{\scriptsize aux}}$ to zero
selects the jammed configurations ($E_{\mbox{\scriptsize
jammed}}=0$), while $T^*$ fixes the energy $E$.

In practice we perform equilibrium MD simulations with a modified
potential energy:

\begin{equation}
U = \frac{T_{\mbox{\scriptsize aux}}}{T^*} E +
E_{\mbox{\scriptsize jammed}},
\end{equation}
and calculate the force on each particle from $\vec{F} = -
\vec{\nabla} U$. Since we need to calculate the force from a
potential energy, only conservative systems can be studied with
this method. Thus we focus our calculations on the system of
frictionless particles. The auxiliary temperature
$T_{\mbox{\scriptsize aux}}$ is controlled by a thermostat which
adjusts the velocities of the particles to a kinetic energy
determined by $T_{\mbox{\scriptsize aux}}$. We start by
equilibrating the system at high temperatures
($T_{\mbox{\scriptsize aux}}$ and $T^*$ $\sim \infty$) and anneal
slowly the value $T_{\mbox{\scriptsize aux}}$ to zero and tune
$T^*$ so as to reach the  value of $E$ that corresponds to the
average deformation energy  obtained during shear.

The partition function is
\begin{equation}
Z = \sum_\nu \exp[- E^\nu/T^* - E^{\nu}_{\mbox{\scriptsize
jammed}}/T_{\mbox{\scriptsize aux}}], \label{partition2}
\end{equation}
from where we obtain the compactivity as
\begin{equation}
T^* = \frac{\partial E}{\partial S} \stackrel
{\mbox{\scriptsize $T_{\mbox{\scriptsize aux}}\to 0$}}{\longrightarrow} X_E,
\end{equation}
Thus at the end of the annealing process ($T_{\mbox{\scriptsize
aux}}\to 0$), $T^*(E)=X_{E}(E)$, since in this limit we are
sampling the configurations with vanishing fraction of moving
particles at a given $E$.

At the end of the protocol the compactivity at a given deformation
energy can be obtained as illustrated in Fig. \ref{annealing}. The
remarkably result
is that  the compactivity and the effective
temperature obtained dynamically are found to coincide
to within the computational error \cite{mk},
\begin{equation}
X_E \approx T_{\mbox{\scriptsize eff}}.
\end{equation}

This suggests
the validity of the effective temperature as a dynamical
estimate of the compactivity, and more importantly, justifies the
use of the novel statistical measurements we have presented in
characterising the macroscopic properties of the system.

\section{Conclusions}

We have experimentally tested the existence of the
effective temperature for sheared granular system
for a given range of particle sizes and densities.
It remains to be seen how robust
our results are under a more extended set of parameters. These include
the use of
particles of different shapes, changing the interstitial liquids, etc.  It
would also be important to test whether the effective temperature of
the packing is the same under different types of driving, for instance
under tapping or shaking, and for different observables.
If the effective temperature is a proper state variable it should
be independent of the observable. Experimentally it would require to measure
the temperature from, for instance, the
volume fluctuations.
In fact an estimation of the compactivity as $X=\partial V / \partial S$ has
been recently obtained in a system of compressed emulsions
observed under the confocal microscope by Bruji\'c {\it et al.} \cite{jasna}.
It is still an open question what is the relation between the compactivity
obtained from volume fluctuations and the effective temperature
obtained from energy considerations.
More experimentation is
needed to fully understand whether the present definition of effective
temperature has a physical thermodynamic meaning.  Our results are a first
step forward in this direction.

Computer simulations shed some light on the thermodynamic meaning
of the effective
temperature. The fact that slow relaxation modes can be
characterized by a temperature raises the question of the
existence of a form of ergodicity for the structural motion,
allowing a construction of a statistical mechanics ensemble for
the slow motion of the grains. This argument leads us back to the
ideas of the thermodynamics of jammed states. In parallel to these
dynamical measurements obtained numerically and experimentally,
the same information is drawn from the
system by a flat statistical average over the jammed
configurations done numerically. Once a sample of
 the static configurations have been
visited by the system, the compactivity $X_E$ can be calculated
from the statistics of the canonical ensemble of the jammed
states. The logarithm of the available configurations at a given
energy and volume reveals the entropy, from which the compactivity
is calculated. Our explicit computation shows that the temperature
arising from the Einstein relation  can be understood
in terms of the configurational compactivity $X_{E}$ arising from
the statistical ensemble of jammed states. This provides
evidence for the validity of the thermodynamic approach.
However, much more experimentation is needed to fully
test the thermodynamics of grains.

In summary, our experimental and numerical
results suggest that the problem of slowly moving
granular matter near jamming could be a nontrivial generalization of
statistical mechanics to this kind of
far-from-equilibrium dissipative system.

\vspace{1cm}

We are deeply grateful
to M. Shattuck for help in the design of the experiments and
J. Kurchan and  J. Bruji\'c for discussions.
We acknowledge financial support from the DOE, Division of Materials
Sciences and Engineering, DE-FE02-03ER46089,
 and the National Science
Foundation, DMR-0239504.

\clearpage

\clearpage

\clearpage

\clearpage

\clearpage

\clearpage


\clearpage

\end{document}